\documentclass[11pt]{article}
\usepackage{moriond,epsfig}

\def\Journal#1#2#3#4{{#1} {\bf #2}, #3 (#4)}

\def\PRL{\em Phys. Rev. Lett.}
\def\PRD{{\em Phys. Rev.} D}

\def\be{\begin{equation}}
\def\ee{\end{equation}}
\def\bea{\begin{eqnarray}}
\def\eea{\end{eqnarray}}

\def\be{\begin{equation}}
\def\ee{\end{equation}}

\def\bi{\begin{itemize}}
\def\ei{\end{itemize}}
\def\bc{\begin{center}}
\def\ec{\end{center}}
\def\and{\/\mbox{and}}

\newcommand{\TeV}{\ensuremath{\mathrm{Te\kern -0.1em V}}}
\newcommand{\GeV}{\ensuremath{\mathrm{Ge\kern -0.1em V}}}
\newcommand{\MeV}{\ensuremath{\mathrm{Me\kern -0.1em V}}}

\newcommand{\GeVCSq}{\ensuremath{\GeV\!/c^2}}
\newcommand{\GeVC}{\ensuremath{\GeV\!/c}}

\newcommand{\zmm}{{ Z}^0\rightarrow{ \mu}^+ { \mu}^-}

\newcommand{\et}{\ensuremath{E_{\rm T}}}
\newcommand{\pt}{\ensuremath{p_{\rm T}}}

\newcommand{\pbinv} {{\mathrm{pb}^{-1}}}

\newcommand{\met}{\mbox{${\hbox{$E$\kern-0.6em\lower-.1ex\hbox{/}}}_{\rm T}\:$}}
\newcommand{\metx}{\mbox{${\hbox{$E$\kern-0.6em\lower-.1ex\hbox{/}}}_T\:$}}
\newcommand{\mety}{\mbox{${\hbox{$E$\kern-0.6em\lower-.1ex\hbox{/}}}_T\:$}}

\newcommand{\bit}{\begin{itemize}}
\newcommand{\eit}{\end{itemize}}
\newcommand{\bce}{\begin{center}}
\newcommand{\beq}{\begin{equation}}
\newcommand{\ece}{\end{center}}
\newcommand{\rg}{\rightarrow} 

\newcommand{\mo}{M_{0}}
\newcommand{\ao}{A_{0}}
\newcommand{\miz}{M_{1/2}}
\newcommand{\tb}{{\rm tan}\beta}
\newcommand{\ntwo}{\tilde{{\chi}}^{0}_{2}}
\newcommand{\cone}{\tilde{{\chi}}^\pm_{1}}
\newcommand{\none}{\tilde{{\chi}}^{0}_{1}}

\newcommand{\ges}{\raisebox{-.7ex}{$\stackrel{\textstyle >}{\sim}$}}

\begin{document}
%

\title{SUPERSYMMETRY RESULTS AT THE TEVATRON}

\author{GIULIA MANCA\footnote{manca@fnal.gov.}\\ (On behalf of the CDF and D\O\ collaborations)}

\address{University of Liverpool, Department of Physics,\\ 
Oxford street, Liverpool L69 7ZE, United Kingdom}

\maketitle\abstracts{
The Run II physics programme of the Tevatron is proceeding 
with more than 300$~\pbinv$ of analysis quality data, collected at a
centre-of-mass energy of 1.96 \TeV. 
Searches for supersymmetric particles are starting to set new
limits, improving over the LEP and Run I results and exploring new regions
of parameter space. We present recent results in Supersymmetry with the
upgraded CDF and D\O\ detectors and give some prospects
for the future of these searches.
}

\section{Introduction}

Supersymmetry(SUSY) is one of the most appealing theories for 
physics beyond the Standard Model(SM), as it solves the 
hierarchy problem and could provide a good candidate
for the cold dark matter(CDM) in the universe.
SUSY predicts the existence of a super-partner
for each SM particle, sharing the same
quantum numbers but differing by half a unit of spin.
Despite numerous attempts, SUSY particles(``sparticles'')
have not yet been observed;
this means that SUSY must be broken and 
sparticles must be heavier than their SM counterparts.
A new quantum number is introduced, R-Parity($R_P$), 
which is $+1$ for SM and $-1$ for SUSY particles.
The conservation of $R_P$ implies that the lightest
supersymmetric particle(LSP) is stable and
escapes the experimental apparatus undetected,
causing a striking experimental signature of large missing transverse
energy($\met$).
Most SUSY theories predict the sparticle spectrum to become
accessible at the TeV scale;
in these scenarios the Tevatron might be able to discover 
SUSY before the advent of the LHC.
We will present searches for various sparticles
at the CDF
and D\O\ experiments; the results are interpreted
in $R_P$ conserving and violating(RPV)  models.

\section{Charginos and Neutralinos}

We present searches for associated production of 
the lightest chargino($\cone$) and 
second to lightest neutralino($\ntwo$),
in three different models.\\
$\bullet$ {\it  mSugra: three leptons plus missing transverse energy.}
The scenario of minimal Sugra(mSugra) is a Grand Unification
Theory which includes Gravity; owing to the small
number of free parameters ($\mo$, $\miz$, $\ao$, $\tb$, $sign(\mu)$),
mSugra is very popular  
in experimental searches. In mSugra 
the lightest neutralino($\none$)
is the LSP and a CDM candidate. 
The LEP2 limit of $M_{\cone} >$ 103.5~\GeVCSq\ implies
the squark($\tilde{q}$) and gluino($\tilde{g}$) 
masses 
to be $\ges 300~\GeVCSq$, meaning that
 strong sparticle production 
at the Tevatron is suppressed.
Thus, the associated \nobreak{production}
\begin{table}[t]
\begin{minipage}[c]{0.48\textwidth} {
\vspace*{-0.0cm} 
\caption{The number of observed events and the SM expectation in 320~$\pbinv$ of
data for the D\O\ trilepton analyses and their sum.}  \label{tab::d0_trilep}
\bce \resizebox{\textwidth}{!} {     
\begin{tabular}{|c c c c|}	     
\hline
{\bf mode} 
& $\mathbf{p_T^{\ell_1},p_T^{\ell_2},p_T^{\ell_3}}$(\GeVC)  
& {\bf SM expected} & {\bf Observed} \\
\hline				     
$ee+\ell$        &12,8,4 & 0.21$\pm$0.12 & 0   \\ 
$e\mu+\ell$      &12,8,7 & 0.31$\pm$0.13 & 0   \\ 
$\mu\mu+\ell$    &11,5,3 & 1.75$\pm$0.57 & 2   \\ 
$\mu^\pm\mu^\pm$ &11,5   & 0.64$\pm$0.38 & 1   \\ 
$e\tau+\ell$     & 8,8,5 & 0.58$\pm$0.14 & 0   \\ 
$\mu\tau+\ell$   &14,7,4 & 0.36$\pm$0.13 & 1   \\ \hline
Total	         & & 3.85$\pm$0.75       & 4   \\ 
\hline				     
\end{tabular} }			     
\ece				     
}
\end{minipage}\hfill
\begin{minipage}[c]{0.5\textwidth} {
of $\cone\ntwo$ would 
be the dominant SUSY production mechanism if 
sparticles are accessible at these energies.
D\O~\cite{d0_np}  has looked  for the ``tri-lepton'' signal
from  $p\bar{p}\rightarrow\cone\ntwo$
followed by $\ntwo\rightarrow\ell\bar{\ell}\none$
and~$\cone\rightarrow\ell\nu\none$.
Events are selected with large $\met$ and 
two isolated leptons
satisfying analysis
 dependent 
topological cuts.
The identification requirements
on the third lepton are loosened
to increase acceptance and include 
hadronic decays  of $\tau$ leptons.
The dominant SM 
}\hfill
\end{minipage}
\vspace{-0.7cm}
\end{table}
\noindent 
background sources are
dibosons and QCD production.
The number of observed events is 
reported in Tab.~\ref{tab::d0_trilep},
together with the lepton selection criteria
and the number of 
SM expected events in each channel.
No evidence of SUSY is observed.
This translates into a limit on 
$\sigma\cdot Br$ 
\begin{figure}[h]
\begin{minipage}[c]{0.46\textwidth} {
\vspace*{-0.2 cm}
of 0.2~pb and on  the mass
of the chargino of 116~\GeVCSq\ for the  scenario with
 $M(\tilde{\ell})\simeq M(\ntwo)$
 and no 
slepton
mixing(``3-l max''~\cite{d0_np}).
The combined limit is shown
 in 
Fig.~\ref{fig::d0_trilep};
the main systematic uncertainties 
on the limit calculation are those 
related to the
statistics of the Monte Carlo background 
samples and the modelling of the QCD background.
This result 
 reaches for the first time 
 beyond the Run~I and LEP2 limit in mSugra 
(for this 
choice of parameter space).
CDF also has preliminary results~\cite{cdf_web} for searches for 
$\cone\ntwo$ in the $ee+\ell$ channel; the results in the other 
channels are imminent.
}
\end{minipage}\hfill
\begin{minipage}[c]{0.5\textwidth} {
\vspace*{-0.3 cm}
\bce
\epsfig{figure=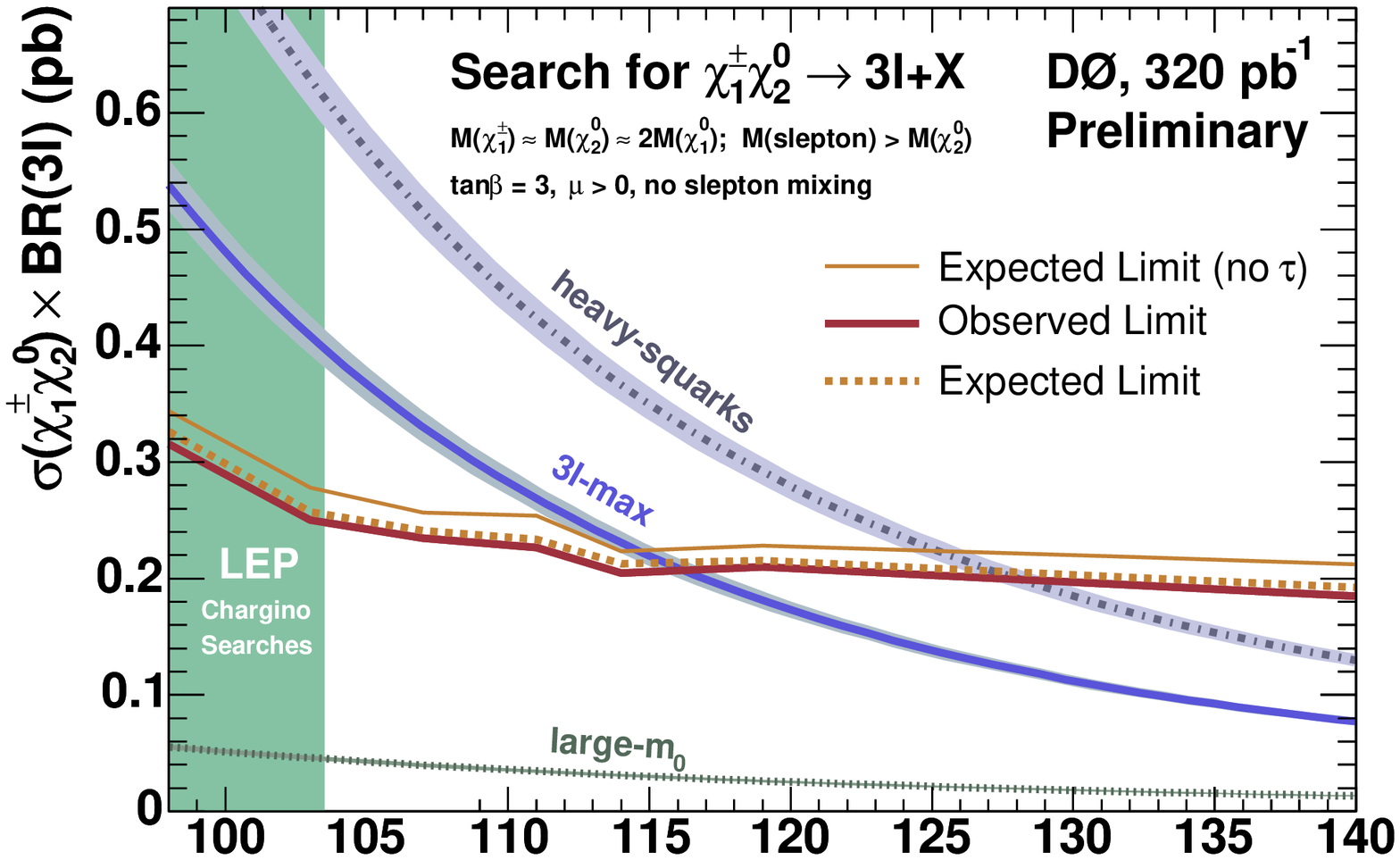,width=\textwidth,height=5.cm}
\vspace*{-1.0cm}\begin{flushright}{\mbox{\small{\bf Chargino Mass (\GeVCSq)}}}\end{flushright}
\ece
\caption{D\O\ combined limit 
 	on the $\sigma\cdot Br$ for 
	$p\bar{p}\rightarrow\cone\ntwo\rightarrow 3\ell+\met$
	as a function of the chargino mass for three different mSugra models.
	}\label{fig::d0_trilep}
}
\end{minipage}\hfill
\end{figure}
\newline$\bullet$ {\it GMSB: diphotons plus missing transverse energy.}
In Gauge-Mediated Supersymmetry Breaking(GMSB) models
a light gravitino is the LSP and 
the next to lightest SUSY particle(NLSP) 
\vspace*{-0.75cm}
\begin{figure}[h]
\begin{minipage}[c]{0.46\textwidth} {
\bce
\epsfig{figure=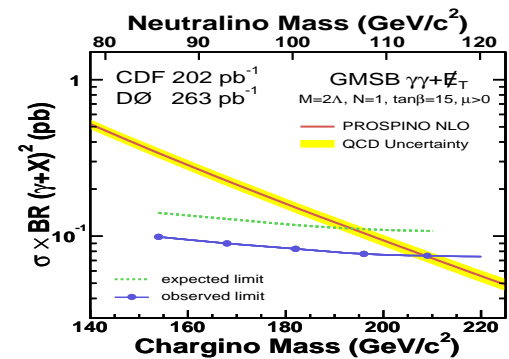,width=\textwidth,height=4.9cm}
\ece
\caption{CDF and D\O\ combined cross section times branching ratio limit 
	for the search for GMSB SUSY events in the $\gamma\gamma\met$ channel.
	}\label{fig::ggmet}
}
\end{minipage}\hfill
\begin{minipage}[c]{0.53\textwidth} {
is
expected to decay in a photon and the LSP if $R_P$ 
is conserved.
The main production mode 
at the Tevatron is 
predicted to be $p\bar{p}\rightarrow\cone\tilde{{\chi}}^\mp_{1}$ or $\cone\ntwo$, where
each gaugino pair cascades down to two $\none$s, leading to a final
state with $\gamma\gamma\met$.
Both CDF and D\O\  have performed
searches in this
channel~\cite{ggmet_run2}.
CDF(D\O) selects events requiring two 
photons above 13(20)~\GeV\
and $\met$ above 45(40)~\GeV. CDF(D\O) observes zero(two) 
events, with  0.3$\pm$0.1(3.7$\pm$0.6) SM events expected.
With these results CDF and D\O\ are able to set a limit 
on 
the mass of the chargino of 167
 and  195~\GeVCSq\ respectively.
The main systematic uncertainty 
comes from 
the photon identification  efficiency in both
}
\end{minipage}\hfill
\vspace*{-0.5 cm}
\end{figure}

\noindent cases.
CDF and  D\O\ combined the two analyses 
obtaining a limit on $M_{\cone}$ of 
209~\GeVCSq 
(see Fig.~\ref{fig::ggmet}), 
significantly improving  over the single experiments and Run~I
 results.\vspace{0.4cm}
\\
$\bullet$ {\it Other models: CHAMPS.}
D\O\ has looked for electrically charged long-lived 
massive particles(CHAMPS) in 390~$\pbinv$ of data.
Using timing information in the muon system, 
events are selected by requiring two isolated muons with 
\pt$>$15~\GeVC\ and 
speed significantly smaller than $c$. 
No events have been observed, with 0.66$\pm$0.06 events expected,
measured in $\zmm$ data.
The main systematic uncertainty comes from 
the $\mu$ efficiencies and 
the time measurement.
D\O\  sets
a limit of 174~\GeVCSq\ on the mass of the stable chargino in 
Anomaly Mediated Supersymmetry Breaking(AMSB) models where the
$\cone$ is the NLSP and a long-lived particle.
This is the most stringent limit to date
in this model. The result has also been interpreted in 
GMSB with the lightest stau NLSP
and in AMSB for long-lived charged higgsinos~\cite{d0_np}.

\section{Squarks and Gluinos}

%
$\bullet$ {\it  mSugra: jets plus missing transverse energy.}
In the most general mSugra models with low $\tb$ the first five squarks
are predicted to be heavy and almost degenerate in mass.
The cross-section for $p\bar{p}\rightarrow\tilde{q}\bar{\tilde{q}}$
should then effectively be the sum of the cross-sections over the ten
squark species and would be large at hadron colliders if 
the squarks are kinematically accessible.
The signature for
$\tilde{q}\tilde{q},\tilde{q}\tilde{g}$ and
$\tilde{g}\tilde{g}$ production and decay is
two to four jets 
(from $\tilde{q}\rightarrow q\tilde{\chi}$ if $M(\tilde{q})>M(\tilde{g})$
and $\tilde{g}\rightarrow qq\tilde{\chi}$ if vice versa) and 
$\met$
coming from the LSPs.
\begin{table}[t]		     
\caption{Number of events and $\tilde{g},\tilde{q}$ mass limits in
the D\O\ squark-gluino searches in 310~$\pbinv$.}  \label{tab::sqgl}
\bce \resizebox{\textwidth}{!} {     
\begin{tabular}{|c c c c c c c|}	     
\hline				     
{\bf mode} 
& $\mathbf{{\it E_T^{j}}}$(\GeV) 
& $\mathrm{\mathbf{\Sigma {\it E_T^j}\,(\GeV)}}$ & $\met$ ($\GeV$) 
& {\bf SM expected} & {\bf Observed}  &  {\bf mass limit($\GeVCSq$)}\\ 
\hline				     
di-jet           &60,50       & 250 & 175 & 12.8$\pm$5.4  & 12   & $M_{\tilde{q}}>$318 \\ 
three jets       &60,40,25    & 325 & 100 &  6.1$\pm$3.1  &  5   & $M_{\tilde{q}}>$333 \\ 
four jets        &60,40,30,25 & 175 &  75 &  7.1$\pm$0.9  & 10   & $ M_{\tilde{g}}>$233 \\ 
\hline				     
\end{tabular} }			     
\ece				     
\end{table}			     
\begin{figure} 
\vspace*{-0.5cm}
\bce			 
\epsfig{figure=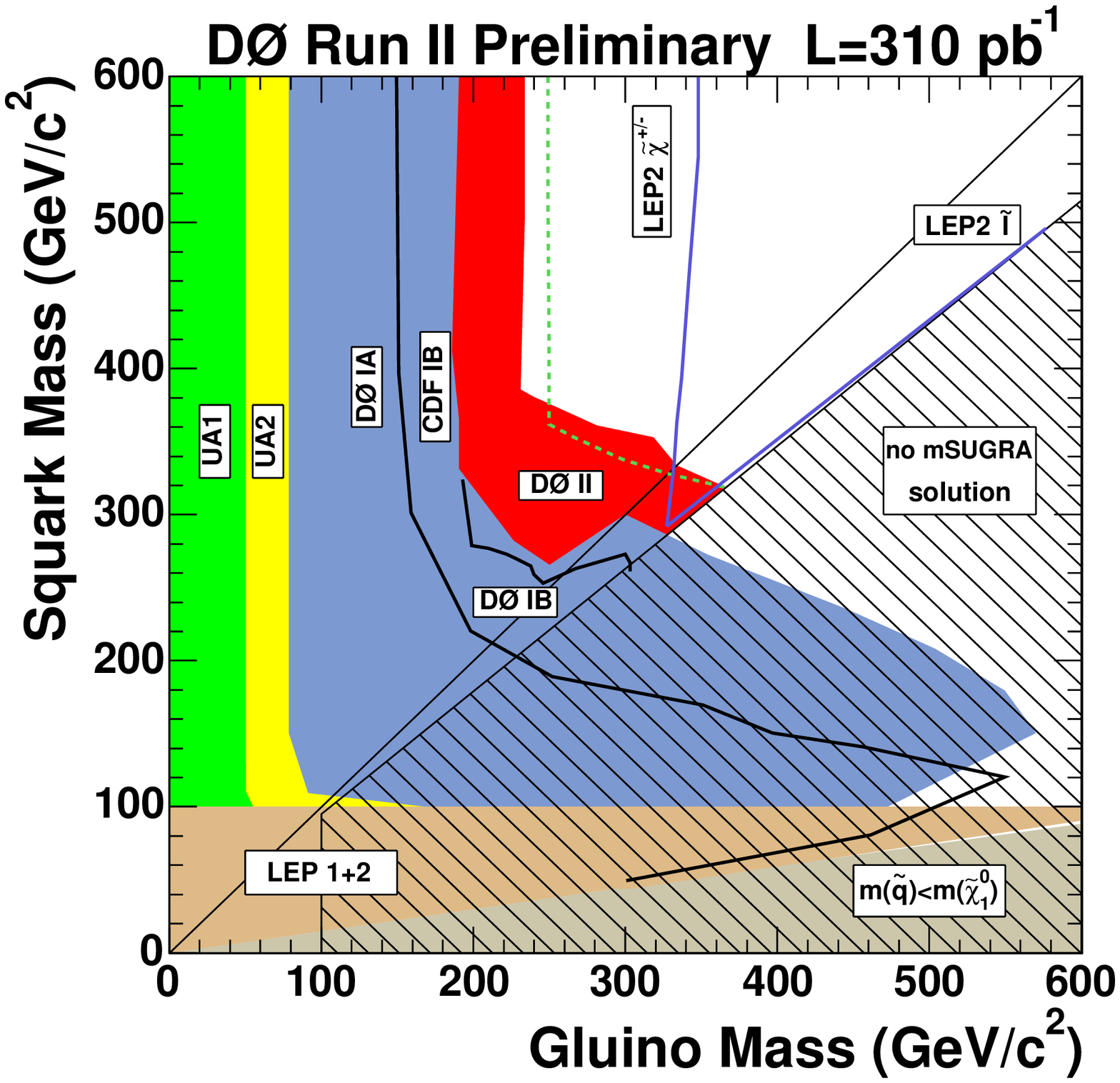,width=0.45\textwidth,height=5cm}
\epsfig{figure=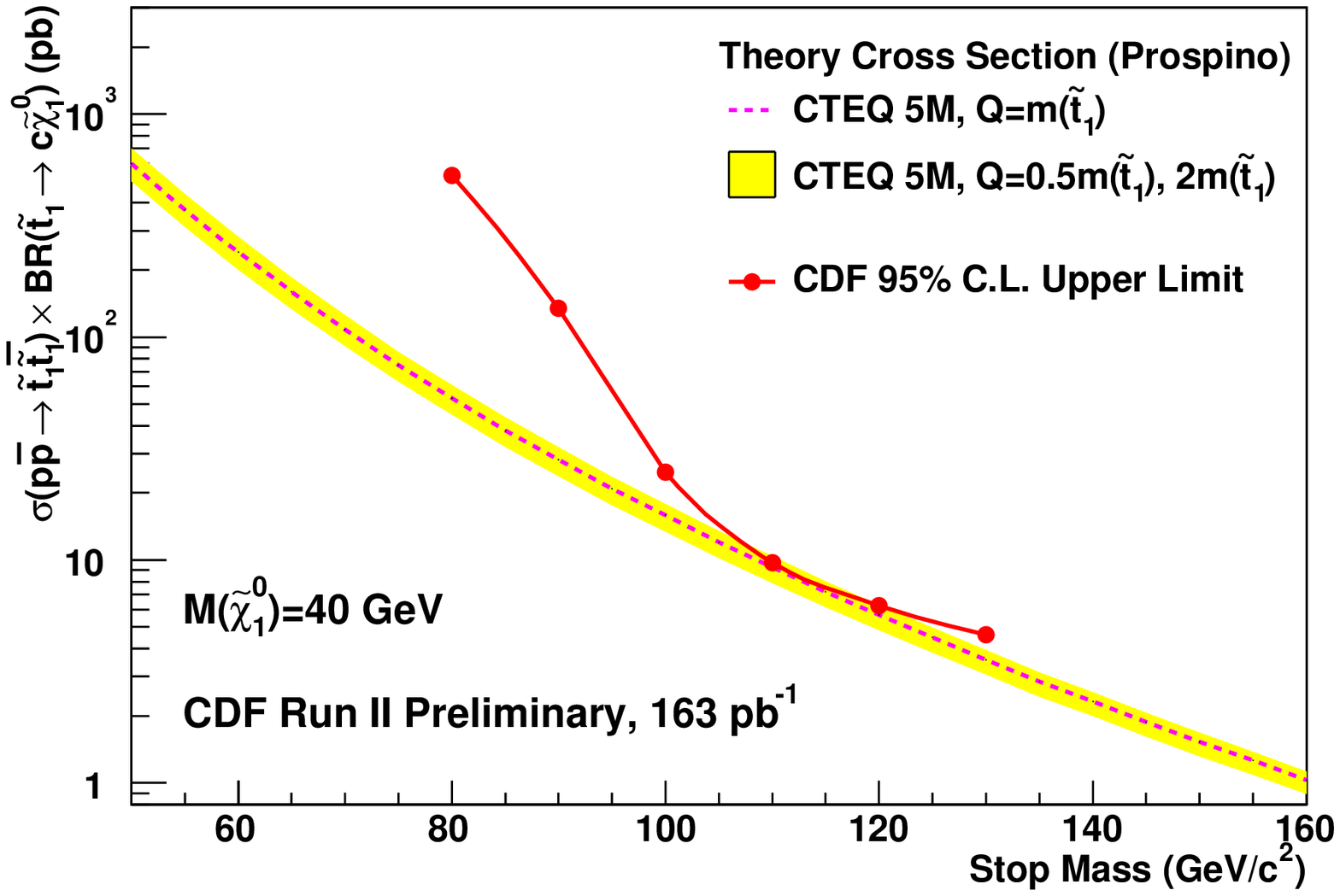,width=0.45\textwidth,height=5cm}
\ece					 
\caption{{\it On the left:} mSugra limits on $M_{\tilde{q}}$ and
$M_{\tilde{q}}$ with $\tb=3,\ao=0,\mu<0$ and $\tilde{q}=\tilde{u},\tilde{d},
\tilde{c},\tilde{s},\tilde{b}$. The red region corresponds to 
the D\O\ analyses.
The di-jet analysis limit
lies above the dashed area in the region of 
$M(\tilde{g})\simeq 350\,\GeVCSq$
and $M(\tilde{q})\simeq 320\,\GeVCSq$. The 3-jets limit is on the
diagonal, for $M(\tilde{q})= M(\tilde{g})\simeq 333\,\GeVCSq$, whilst the
4-jets limit is on the right of the dashed line (corresponding
to the expected limit) at $M(\tilde{g})\simeq 230~\GeVCSq$.
{\it On the right:} CDF 95\% C.L. on $\sigma(p\bar{p}\rightarrow \tilde{t}_1\bar{\tilde{t}}_1 \rightarrow(c\none)(\bar{c}\none))$ as a function of $M(\tilde{t}_1)$. 
}\label{fig::sqgl}			 
\end{figure} 				 
D\O\ has looked for squark and gluinos in 310~$\pbinv$
of data with three analyses, optimised to 
search for $\tilde{q}\bar{\tilde{q}}$, $\tilde{q}\tilde{g}$ 
or $\tilde{g}\tilde{g}$ assuming 
 $M(\tilde{q})<M(\tilde{g})$,  
 $M(\tilde{q})\simeq M(\tilde{g})$ or 
 $M(\tilde{q})>M(\tilde{g})$ 
respectively.
Table~\ref{tab::sqgl} shows the selection requirements,
the SM expected and the observed
events for the three cases. 
The main backgrounds are 
$Z(\rg\nu\bar{\nu})+jets$,
$W(\rg\tau\nu)+jets$ and $t\bar{t}\rightarrow b\bar{b}jj\ell\nu$
for two, three and four jets respectively, while
the main systematic comes from the jet energy scale.
No evidence of SUSY
is observed, which leads to a limit on $M(\tilde{q},\tilde{g})$
(see Table~\ref{tab::sqgl}
and Fig.~\ref{fig::sqgl}, left).
CDF has also recently obtained  results in this channel,
which are compatible with D\O.
%
\\$\bullet$ {\it  Stop searches.}
%
CDF has searched~\cite{cdf_web} for 
$\tilde{t}_1\bar{\tilde{t}}_1$ pair production 
assuming $Br(\tilde{t}_1\rightarrow c\none)$=1
and $M_{\none(LSP)}>40~\GeVCSq$.
The decay $\tilde{t}_1\rightarrow c\none$
dominates via a one-loop diagram in the absence 
of flavour changing neutral currents if
$M_{\tilde{t}_1}<M_{b}+M_{\cone}$, 
$M_{\tilde{t}_1}<M_W+M_{b}+M_{\none}$,
$M_{\tilde{t}_1}<M_{b}+M_{\tilde{\nu}}$ and
$M_{\tilde{t}_1}<M_{b}+M_{\tilde{\ell}}$.
The signature for this process is a pair
of acollinear heavy flavour jets in the transverse plane,
large $\met$ and no isolated high \pt\ leptons. 
The existing CDF(D\O) Run~I limit is
$M_{\tilde{t}_1}>119(122)~\GeVCSq$  for $M_{\none}$
up to 40(45)~$\GeVCSq$.
CDF~\cite{cdf_web} has looked for this signature, 
with and without the requirement of the jets to be tagged by the Silicon 
Vertex Detector.
In 163~$\pbinv$ of data CDF expects 105$\pm$12 SM events 
in the ``pre-tag''sample and 8.3$\pm$2.3 in the ``tag''
sample, and observes 119 and 11 respectively, with 
background dominantly QCD multi-jet  
and $W/Z$+jets production.
The main systematic uncertainty 
comes from the jet energy scale.
Fig.~\ref{fig::sqgl}(right) shows the 95\% confidence limit on the $\sigma\cdot Br$.
This result does not improve the Run~I limit yet.

\section{R-Parity Violation}\label{sec::rpv}

%
%
Three analyses have been performed~\cite{d0_np} by D\O, looking for RPV 
in the decays 
$\none\rightarrow\mu e\nu_e, e e\nu_\mu ,$\\ 
$\none\rightarrow\mu\mu\nu_e, \mu e \nu_\mu $ and 
$\none\rightarrow e \tau\nu_\tau, \tau\tau \nu_e$. 
Here $R_P$ is assumed to be conserved in the production, with the
main process being $p\bar{p}\rightarrow\cone\ntwo$.
The signature for 
these analyses is four charged leptons and missing energy coming 
from the neutrinos.
The events have  been selected with at least three leptons 
 and missing
energy.  No evidence of RPV SUSY is observed,
which translates in a limit on the masses of the $\none$ and the 
$\cone$ (see Table~\ref{tab::rpv}).\\
D\O\ has also searched~\cite{d0_np} for the RPV process
$u\bar{d}\rightarrow\tilde{\mu}$,
with 
$\tilde{\mu}\rightarrow\none\mu
\rightarrow (\mu\bar{\tilde{\mu}^*}) \mu$, 
where the virtual smuon decays through
$\tilde{\mu} \rightarrow u\bar{d}$
again violating $R_P$.
The same RPV coupling $\lambda^\prime_{211}$ is involved
in both vertices; all the other couplings are
assumed to be zero.
Events are selected requiring two jets with \et$>15\,\GeV$
and two high \pt\ isolated muons.
The number of events and the corresponding limit are summarised
in Table~\ref{tab::rpv}.\\
%
%
%
%
CDF has performed a search~\cite{cdf_web} 
for $\tilde{t}_1\bar{\tilde{t}}_1$
production 
followed by $R_P$ violating 
decay of the stop into $b\tau$
with $Br(\tilde{t}_1 \rightarrow b\tau)$=1.
The signature for this analysis is 
either an electron or a muon (from 
$\tau\rightarrow\ell\bar{\nu_\ell}\nu_\tau$),
a hadronically decaying tau and at least 
two jets.  
%
CDF expects 2.6$\pm$0.6 $e\tau$ and 2.2$\pm$0.5 $\mu\tau$
events 
from SM processes, and observes 2 and 3 
respectively. 
The dominant uncertainty on the mass limit comes from the 
PDFs(10\%) and $\met$ estimate(3\%).
As the data are in good agreement with the
expectation, a limit of 129~$\GeVCSq$
is set on the mass of the stop
as shown in Table~\ref{tab::rpv}.
\begin{table}[t]		     
\caption{The number of SM expected and observed events for the 
RPV analyses described in section~\ref{sec::rpv}.}  \label{tab::rpv}
\vspace{0.2cm}		     
\bce \resizebox{0.9\textwidth}{1cm} {     
\begin{tabular}{|c c c c c c c|}	     
\hline				     
{\bf Experiment} & {\bf RPV process} & {\bf RPV coupling} 
& $\mathcal{L}(\pbinv)$ 
& {\bf SM expected} & {\bf Observed} &  {\bf mass limit(\GeVCSq)}\\
\hline
D\O\ &$\none\rightarrow\mu\mu\nu_e, \mu e \nu_\mu$ &$\lambda_{122}$& 160 & 0.6$\pm$1.9 & 2  & $M({\cone})>$165,$M({\none})>$84\\ 
D\O\ &$\none\rightarrow\mu e\nu_e, e e\nu_\mu$     &$\lambda_{121}$& 238 & 0.5$\pm$0.4 & 0  & $M({\cone})>$181,$M({\none})>$95\\ 
D\O\ &$\none\rightarrow e \tau\nu_\tau, \tau\tau \nu_e$&$\lambda_{133}$	& 200 & 1.0$\pm$1.4 & 0  & $M({\cone})>$118,$M({\none})>$66\\ 
\hline
D\O\ &$\tilde{\mu}\rightarrow\none\mu$, $\tilde{\mu} \rightarrow u\bar{d}$  
	&$\lambda^\prime_{211}$=0.07 & 154 & 1.1$\pm$0.4 & 2  &$M({\tilde{\mu}})>$255,$M({\none})\simeq$100\\		     
CDF  &$\tilde{t}_1 \rightarrow b\tau$  &$\lambda^\prime_{333}$	  & 200 & 4.8$\pm$0.7 & 5  &$M({\tilde{t}_1})>$129\\ 
\hline
\end{tabular} }			     
\ece		
\vspace*{-0.3cm}		     
\end{table}			     
This result can also be interpreted as a limit on the third 
generation leptoquark ($LQ_3$), assuming $Br(LQ_3\rightarrow \tau b$)=1.

\section{Conclusions and prospects}

The CDF and D\O\ detectors are running efficiently 
and have already collected more than three times the luminosity
of Run~I.
Most of the SUSY limits we have presented 
are an improvement 
over the Run~I results, and are
probing the region beyond the LEP2 limits. 
Many more results are to come
with the present and future data; the exciting era has just begun.

\end{document}